\documentstyle[11pt,epsfig,amsbsy]{article}

\def\be{\begin{equation}}
\def\ee{\end{equation}}
\def\ba{\begin{eqnarray}}
\def\ea{\end{eqnarray}}
\def\bs{\boldsymbol}
\def\ga{\mathrel{\raise.3ex\hbox{$>$\kern-.75em\lower1ex\hbox{$\sim$}}}}
\def\la{\mathrel{\raise.3ex\hbox{$<$\kern-.75em\lower1ex\hbox{$\sim$}}}}

\newcommand{\fr}[2]{\frac{#1}{#2}}

\newcommand{\ode}{\omega_{\rm{DE}}}
\newcommand{\odeo}{\omega_{\rm{DE}}^{(0)}}
\newcommand{\Ode}{\Omega_{\rm{DE}}^{(0)}}

\newcommand{\AST}{astro-ph}
\newcommand{\ApJ}{Astrophys.\ J.\ }
\begin{document}

\baselineskip=16pt
\begin{titlepage}
\begin{center}

\vspace{0.5cm}

\large {\bf Can Strong Gravitational Lensing Constrain Dark
Energy?} \vspace*{5mm} \normalsize

{\bf Seokcheon Lee$^{\,1}$} and {\bf Kin-Wang Ng$^{\,1,2}$}

\smallskip
\medskip

$^1${\it Institute of Physics,\\ Academia Sinica,
 Taipei, Taiwan 11529, R.O.C.}

$^2${\it Institute of Astronomy and Astrophysics,\\
 Academia Sinica, Taipei, Taiwan 11529, R.O.C.}

\smallskip
\end{center}

\vskip0.6in

\centerline{\large\bf Abstract}

We discuss the ratio of the angular diameter distances from the
source to the lens, $D_{ds}$, and to the observer at present,
$D_{s}$, for various dark energy models. It is well known that the
difference of $D_s$s between the models is apparent and this
quantity is used for the analysis of Type Ia supernovae. However
we investigate the difference between the ratio of the angular
diameter distances for a cosmological constant,
$(D_{ds}/D_{s})^{\Lambda}$ and that for other dark energy models,
$(D_{ds}/D_{s})^{\rm{other}}$ in this paper. It has been known
that there is lens model degeneracy in using strong gravitational
lensing. Thus, we investigate the model independent observable
quantity, Einstein radius ($\theta_E$), which is proportional to
both $D_{ds}/D_s$ and velocity dispersion squared, $\sigma_v^2$.
$D_{ds}/D_s$ values depend on the parameters of each dark energy
model individually. However, $(D_{ds}/D_s)^{\Lambda} -
(D_{ds}/D_{s})^{\rm{other}}$ for the various dark energy models,
is well within the error of $\sigma_v$ for most of the parameter
spaces of the dark energy models. Thus, a single strong
gravitational lensing by use of the Einstein radius may not be a
proper method to investigate the property of dark energy. However,
better understanding to the mass profile of clusters in the future
or other methods related to arc statistics rather than the
distances may be used for constraints on dark energy.

\vspace*{2mm}

\end{titlepage}

\section{Introduction}
\setcounter{equation}{0}
Recent observations of high redshift Type Ia supernovae (SNe Ia)
suggested that the expansion of the Universe is currently
accelerating~\cite{SCP}. The cosmic microwave background (CMB)
anisotropy data, indicating a spatially flat universe \cite{CMB}
containing a low value for the cold dark matter (CDM) density
parameter \cite{CDM}, has confirmed that the Universe is
dominantly made up of a component with negative pressure (dark
energy) to make up the critical density today.

The cosmological constant and/or a quintessence field are the most
commonly accepted candidates for dark energy. Although the
cosmological constant is simple and favored by current
cosmological observations, there is $50$ to $120$ orders of
magnitude discrepancy between theory and the measured value
\cite{Weinberg}. The quintessence, which might alleviate this
problem is a dynamical scalar field leading to a time dependent
equation of state parameter~\cite{quint}. Various scalar field
potentials for the quintessence have been
investigated~\cite{SLee2}.

It is important to use various ways of checking for the existence
of dark energy in addition to SNe Ia and CMB anisotropy
constraints on dark energy. A number of other tests have been
considered including ``geometric'' tests using standard
cosmological methods (the galaxy cluster gas mass
fraction~\cite{gasmass}, the location of CMB peaks~\cite{CMBpeaks,
SLee}, the redshift-angular size~\cite{angular}, the strong
gravitational lensing~\cite{SGL} - \cite{Yamamoto}, fluctuations
of the luminosity distance~\cite{RD}, etc.).

The statistics of gravitational lensing of quasars (QSOs) by
intervening galaxies can constrain on the cosmological constant
\cite{Kochanek}. Lensed images of distant galaxies in cluster,
arcs or rings, may provide a bound on the equation of state
parameter of dark energy~\cite{Futamase}. While SNeIa is used to
determine the luminosity distance itself, a gravitational lensing
system can be used measure the ratio of angular diameter
distances. Thus, the gravitational lensing system is regarded as
an independent tool that complements SNe Ia as a probe of dark
energy~\cite{Yamamoto}.

However, the lensing observations primarily depend on the
parameters of lens models with minor dependence on cosmological
parameters~\cite{Chiba}. There is the lens model degeneracy in
both the projected mass density profile and the circular velocity
profile. It is shown that we need to measure the Einstein radius
and the velocity dispersion within ${\cal O} (1)$\% accuracy in
order to put a constraint on $\omega_{DE}$.

In the gravitational lensing, one of the observable quantities
without having any model dependence is the Einstein radius
($\theta_E$), which is proportional to the velocity dispersion
squared ($\sigma_v^2$) and the ratio of the angular distances
$D_{ds}/D_s$, where $D_{ds}$ is the distance from the lens to the
source and $D_s$ is that from the source to the observer. With
different values of cosmological parameters, we can have different
values of $D_{ds}/D{s}$, {\it i.e.} different values of
$\theta_{E}$. Thus, it might be used for probing the property of
dark energy, $\omega_{DE}$. However, there is an ambiguity in
measuring $\sigma_v$. If the error of $\sigma_v$ measurement is
not within the differences of $D_{ds}/D{s}$ between different
cosmological models, then we cannot distinguish the differences
between models by measuring $\theta_E$.

This paper is organized as follows. In the next section we review
the gravitational lensing system with the basic equations used in
lensing observations. We also briefly mention the most popular
lens models. We review the various aspects of errors in modeling
lens in Sec.~$3$. In Sec.~$4$, we check both the differences of
$D_s$ and the differences of $D_{ds}/D{s}$ between the
cosmological constant and other dark energy models. Our conclusion
is in the last section.

\begin{center}
\begin{figure}
\vspace{1cm} \epsfxsize=5.5cm
\centerline{\psfig{file=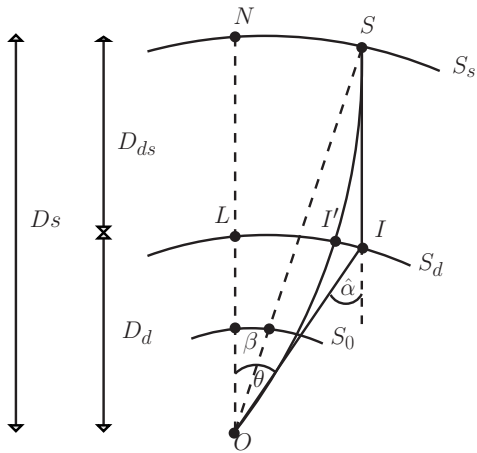, height=6cm}} \caption{A
general lensing system. $L$ is the center of the lens, and the
line through $L$ and the observer $O$ is the optical axis.
$\bs{\beta}$ is an unperturbed angular position of the source
relative to that. $\hat{\bs{\alpha}}$ is the deflection angle of a
light ray, thus an image of the source is observed at position
$\bs{\theta}$. However all angles are very small, we can replace
the real light ray ($SI'O$) by its approximation $SIO$.}
\label{fig:GL}
\end{figure}
\end{center}

\section{Gravitational Lensing and Isothermal Galaxy Models}
\setcounter{equation}{0}

Figure~\ref{fig:GL} shows a simple lensing system
\cite{Schneider}. Consider the source sphere $S_s$, {\it i.e.} a
sphere with radius $D_s$, centered at the observer $O$ and the
deflector sphere $S_d$ with radius $D_d$, {\it i.e.} the distance
to the center of the lens $L$. In addition, consider the observer
sphere $S_o$ where the source would have angular position
$\boldsymbol{\beta}$ (bold faced characters represent
$2$-dimensional vectors in this manuscript) if the light rays from
the source $S$ were not influenced by the gravitational field of
the deflector. However, since light rays are bent by the lens, the
straight line $SO$ is no longer a physical ray path. Rather, there
are light rays which connect the source and the observer but they
are curved near $S_d$. One such ray $SI'O$ is drawn, together with
its approximation $SIO$, consisting of the two asymptotes of the
real ray. The separation of the light ray from the optical axis,
$LI$, is defined as the impact vector $\boldsymbol{\xi}$ in the
lens plane. The angle $\hat{\boldsymbol{\alpha}}$ between the two
asymptotes $SI$ and $IO$ is the deflection angle caused by the
matter distribution $L$, \be \hat{\boldsymbol{\alpha}} = 4G
\int_{R^{2}} \Sigma(\boldsymbol{\xi'})
\fr{\bs{\xi}-\bs{\xi'}}{|\bs{\xi}-\bs{\xi'}|^2} d^2 \bs{\xi'}
\label{alphahat} \ee where the integral is over the lens plane and
$\Sigma(\bs{\xi'})$ is the surface matter density at position
$\bs{\xi'}$ resulting from the projection of the volume mass
distribution of the deflector onto the lens plane. This is valid
when the gravitational field is weak, hence the deflection angle
is small. The observer will thus see the source at the position
$\bs{\theta}$ on his sphere $S_o$.

From the geometry of Fig.\ref{fig:GL}, we can easily derive a
relation between the source position described by the unlensed
position angle $\bs{\beta}$ and the position of  the images
$\bs{\theta} = \bs{\xi} / D_d$ of the source \be \bs{\beta} =
\bs{\theta} - \fr{D_{ds}}{D_s} \hat{\bs{\alpha}} (\bs{\xi})
\label{lenseq1} \, . \ee We can reexpress this equation by using
the distance $\bs{\eta} = D_s \bs{\beta}$ from the source to the
optical axis as \be \bs{\eta} = \fr{D_s}{D_d} \bs{\xi} - D_{ds}
\hat{\bs{\alpha}} (\bs{\xi}) \label{lenseq2} \, . \ee

It is useful to rewrite lens equations (\ref{lenseq1}) and
(\ref{lenseq2}) in dimensionless form by introducing a length
scale $\xi_0$ in the lens plane, which is called as the Einstein
radius in the lens plane and a corresponding length scale $\eta_0
= \xi_0 D_s/D_d$ in the source plane. The Einstein radius in the
lens plane is given by \be \xi_0 = \sqrt{4 G M_d} \cdot
\sqrt{\fr{D_d D_{ds}}{D_s}} \label{Eradius} \, , \ee where $M_d$
is the mass of the lensing object. By use of the definition of the
dimensionless vectors $\bs{x} \equiv \bs{\xi} / \xi_0$ and $\bs{y}
\equiv \bs{\eta} / \eta_0$ as well as the dimensionless surface
mass density (convergence) $\kappa(\bs{x}) = \Sigma(\xi_0 \bs{x})
/ \Sigma_{cr}$ where the critical surface matter density
$\Sigma_{cr} = \fr{1}{4 \pi G} \fr{D_s}{D_d D_{ds}}$, the lens
equations are rewritten as \be \bs{y} = \bs{x} -
\bs{\alpha}(\bs{x}) = \bs{x} - \fr{m(\bs{x})}{\bs{x}}
\label{lenseq3} \, , \ee where $\alpha$ is the scaled deflection
angle, \be \bs{\alpha}(\bs{x}) = \fr{D_d D_{ds}}{D_s \xi_0}
\hat{\bs{\alpha}}(\xi_0 \bs{x}) = \fr{1}{\pi} \int_{R^2} d^2
\bs{x'} \kappa(\bs{x'}) \fr{\bs{x}-\bs{x'}}{|\bs{x}-\bs{x'}|^2} \,
, \label{alpha} \ee and $m(\bs{x})$ is defined as \be m(\bs{x}) =
2 \int_{0}^{x} \kappa(\bs{x'}) \bs{x'} d \bs{x'} \, . \label{mx}
\ee If the lensing object is a point mass, then the cross section
$\sigma$ for strong lensing events is given by \cite{TOG} \be
\sigma = \pi \xi_0^2 \, . \label{sigma} \ee


We need to specify the mass distribution $\Sigma$ to solve the
lens equation (\ref{lenseq3}). One simple and analytic solution of
a differential equation for the radial mass distribution is the
singular isothermal sphere (SIS) \cite{TOG}. Its surface mass
density, projected onto the lens plane, is given by \be
\Sigma(\bs{\xi}) = \fr{\sigma_{v}^2}{2G} \fr{1}{\xi}
\label{SISSigma} \, , \ee where $\sigma_v$ is the velocity
dispersion along the line of sight and the length scale $\xi_0 = 4
\pi \sigma_v^2 \fr{D_d D_{ds}}{D_s}$. With these we get the lens
equation \be \bs{y} = \bs{x} - \fr{\bs{x}}{|\bs{x}|}
\label{SISlenseq}\, , \ee with the convergence \be \kappa(x) =
\fr{1}{2x} \, . \label{SISkappa} \ee Due to its simplicity and the
consistency with the matter distribution of galaxies, the SIS is
frequently used as the gravitational lens model. In addition to
SIS, a nonsingular isothermal sphere (NIS), the singular
isothermal ellipsoid (SIE) and Navarro-Frenk-White
(NFW)~\cite{NFW} are the most commonly used gravitational lens
models. All of them are distinguished by their own surface mass
densities as in the equation (\ref{SISSigma}).

\section{Errors in Modeling Lens}
\setcounter{equation}{0}

In this section we review systematic errors in the ratio
$D_{ds}/D_{s}$ \cite{Yamamoto}. As we mentioned in the previous
section, the various isothermal galaxy models are specified by
their surface mass densities ($\Sigma$), which are the function of
the velocity dispersion ($\sigma_v$) along the line of sight.
Isothermal ellipsoid models (SIE, NIE) have an additional
necessary measurement, the ratio $f$ of the minor axis to the
major axis, $\zeta = \sqrt{\xi_1^2 + f^2 \xi_2^2}$, which is
related to the ellipticity $\epsilon$ by \be f =
\sqrt{(1+\epsilon)/(1-\epsilon)} \, . \label{f} \ee The lens
equation gives an elliptical image of the Einstein ring with the
minor and major axes \be \theta_{\pm} = \theta_{E} \sqrt{1 \pm
\epsilon} \, , \label{thetapm} \ee where \be \theta_{E} = 4 \pi
\sigma_{v}^2 \fr{D_{ds}}{D_s} \, , \label{thetaE} \ee as given in
Eq. (\ref{Eradius}). Thus $D_{ds} / D_{s}$ can be determined by
measuring $\sigma_v$, $f$, and $\theta_{E}$.

Even though gravitational lensing has been used as a useful
cosmological tool to probe the high redshift universe, there are
several problems. Observational quantities depend on a lens model,
which has inherent uncertainties in itself~\cite{errors}. In most
cases, we do not know the property of the lensing object in
detail. Also, the light propagates through the local inhomogeneous
spacetime which deviates from the smoothed Robertson-Walker
metric. Thus, even though the global parameters such as the energy
density parameters are fixed, the distance formula is not uniquely
determined~\cite{DR}. In spite of all these ambiguities, if dark
energy dominates over the mass density, then the optical depth
increases dramatically and its effect is much larger than the
uncertainty arising from the problems in the
formulation~\cite{Fukugita}. However in the following section we
investigate the model independent observable quantity, Einstein
radius.

\section{Angular Diameter Distances}
\setcounter{equation}{0}

In this section, we investigate the ratio of the angular diameter
distances from the lens to the sources ($D_{ds}$) and those from
the source to the present observer ($D_s$) for various dark energy
models. We consider a spatially flat, homogeneous, and isotropic
universe with radiation, matter, and dark energy.

The angular diameter distance from the observer at present to the
source, $D_{s}(0,z_s)$, is defined as \ba D_{s}(0,z_s) &=&
\fr{d(0,z_s)}{\delta} = d_L (1+z_s)^{-2} = a_s \, r_s = \fr{D(0,z_s)}{(1+z_s)} \nonumber \\
&=& \fr{H_{0}^{-1}}{(1+z_s)} \int_{0}^{z_s}
\fr{dz}{\sqrt{\Omega_{m}^{(0)} (1+z)^3 \Bigl(1 +
\fr{(1+z)}{(1+z_{eq})} \Bigr) + \Omega_{\rm{DE}}^{(0)} e^{3
\int_{0}^{z} (1 + \omega_{\rm{DE}}) d \ln (1+z')}}} \, ,
\label{dA} \ea where $d_L$ is the luminosity distance and $z_{eq}$
is the redshift when radiation and matter densities are equal. In
the following numerical calculations, we use $a_{eq} = (1 +
z_{eq})^{-1} = 1/5510$. The angular diameter distance depends on
the present values of energy density contrast of matter
($\Omega_{m}^{(0)}$) and dark energy ($\Omega_{\rm{DE}}^{(0)}$)
and the equation of state of dark energy ($\omega_{\rm{DE}}$) as
in Eqs. (\ref{dA}) and (\ref{dA2}). In a flat universe, we have
the constrain $\Omega_{m}^{(0)} + \Omega_{\rm{DE}}^{(0)} \simeq 1$
when we use the fact that $\Omega_{r}^{(0)} \ll \Omega_{m}^{(0)}
\, , \Omega_{\rm{DE}}^{(0)}$. Especially, dark energy is the
cosmological constant ($\Lambda$) when $\omega_{\rm{DE}}^{(0)} =
-1.0$ and we call the model including cold dark matter and
cosmological constant as $\Lambda$CDM model.

However $H_0D_s$, itself may not be an interesting quantity in the
study of gravitational lensing. Instead, we need to know the
combination of angular diameter distances. We want to investigate
the least model dependent observable quantity for our study and
Einstein radius is one of the proper objects for this purpose. We
can represent the angular diameter distance from the lens to the
source as \be D_{ds}(z_d,z_s) = \fr{H_{0}^{-1}}{(1+z_s)}
\int_{z_d}^{z_s} \fr{dz}{\sqrt{\Omega_{m}^{(0)} (1+z)^3 \Bigl(1 +
\fr{(1+z)}{(1+z_{eq})} \Bigr) + \Omega_{\rm{DE}}^{(0)} e^{3
\int_{0}^{z} (1 + \omega_{\rm{DE}}) d \ln (1+z')}}} \, ,
\label{dA2} \ee where we assume $z_s > z_d$.

\begin{center}
\begin{figure}
\vspace{2cm} \centerline{ \psfig{file=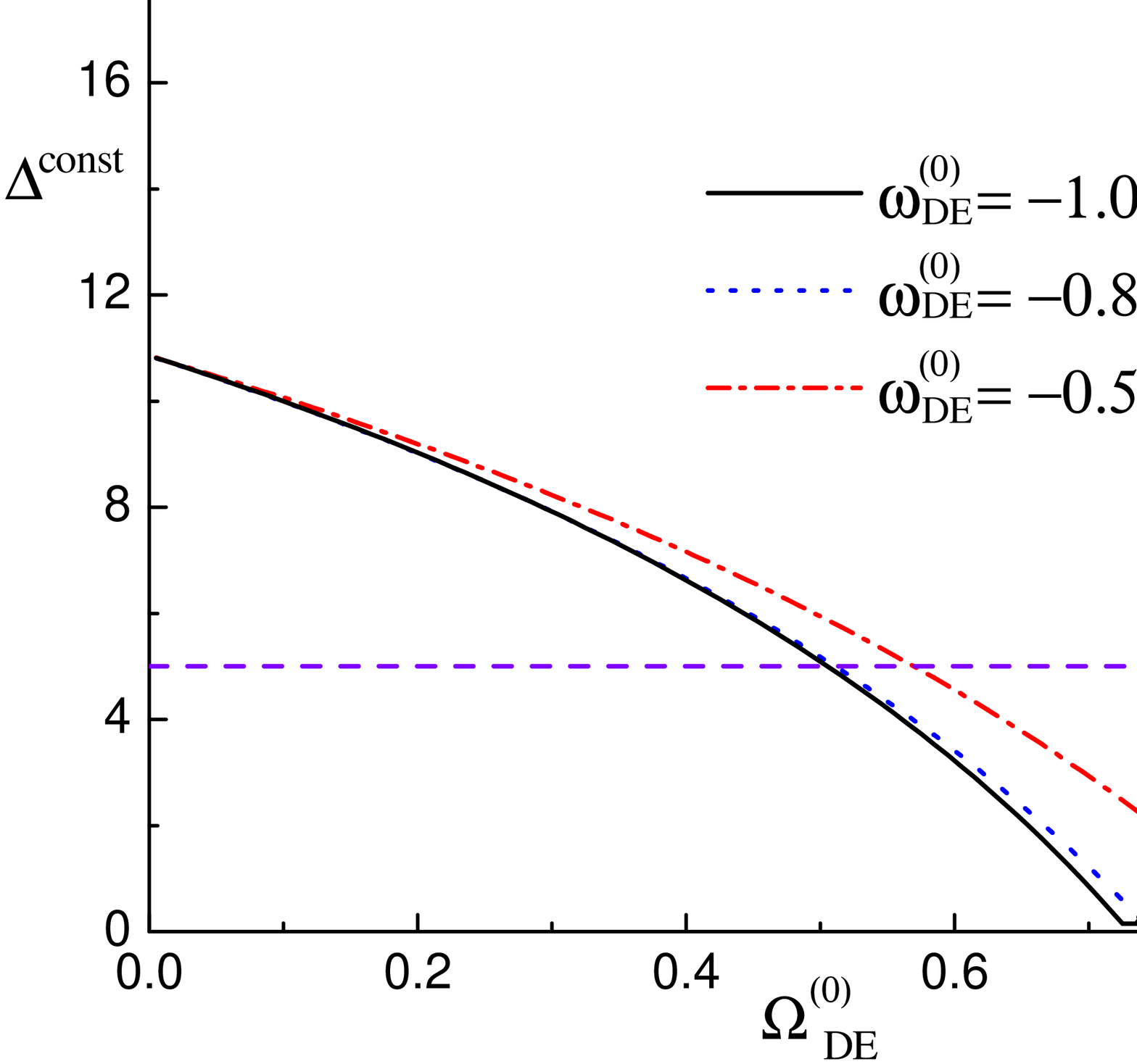,
width=8cm}\psfig{file=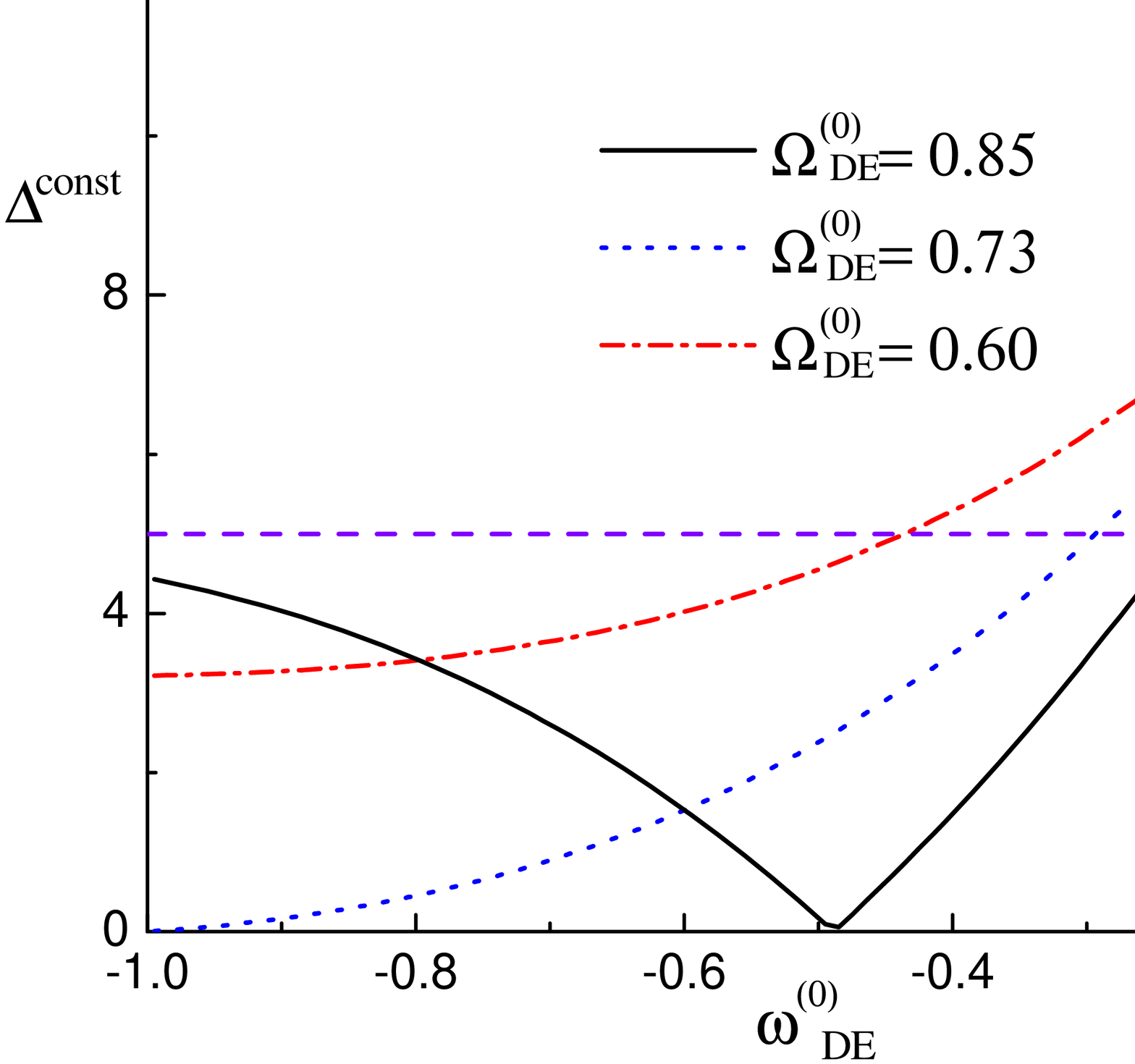, width=8cm} } \vspace{-2cm}
\caption{The ratio of angular diameter distances
$\Delta^{\rm{const}}$ for the constant equation of state of dark
energy ($\omega_{\rm{DE}}$). (a) Dependence on the present value
of dark energy density contrast $\Omega_{\rm{DE}}^{(0)}$ of
$\Delta^{\rm{const}}$ for different present values of equation of
state of dark energy  $\omega_{\rm{DE}}^{(0)}$, : $-1.0$ (solid
line), $-0.8$ (dotted line), and $-0.5$ (dash-dotted line),
respectively. (b) Dependence on the value of
$\omega_{\rm{DE}}^{(0)}$ of $\Delta^{\rm{const}}$ for different
values of $\Omega_{\rm{DE}}^{(0)}$ : $0.85$ (solid line), $0.73$
(dotted line), and $0.60$ (dash-dotted line), respectively.}
\label{fig:ErrR} \vspace{1cm}
\end{figure}
\end{center}

\vspace{-1.0cm}

If we rewrite the Einstein radius, Eq.~(\ref{thetaE}), then the
ratio of the angular diameter distances $D_{ds}/D_{s}$ is given by
\ba R_{ds} &\equiv& \fr{D_{ds}}{D_{s}} = \fr{\theta_{E}}{4 \pi
\sigma_{v}^2} \, , \nonumber \\ &=& \fr{\int_{z_d}^{z_s}
\fr{dz}{\sqrt{\Omega_{m}^{(0)} (1+z)^3 (1 + \fr{(1+z)}{(1+z_{eq})}
) + \Omega_{\rm{DE}}^{(0)} \exp (3 \int_{0}^{z} (1 +
\omega_{\rm{DE}}) d \ln (1+z'))}}}{ \int_{0}^{z_s}
\fr{dz}{\sqrt{\Omega_{m}^{(0)} (1+z)^3 (1 + \fr{(1+z)}{(1+z_{eq})}
) + \Omega_{\rm{DE}}^{(0)} \exp(3 \int_{0}^{z} (1 +
\omega_{\rm{DE}}) d \ln (1+z'))}}} \, . \label{Rds} \ea Thus, the
error that contributes to this ratio due to the error from the
velocity dispersion is \be \Biggl| \fr{\Delta R_{ds}}{R_{ds}}
\Biggr| = 2 \fr{\Delta \sigma_{v}}{\sigma_{v}} \, .
\label{DeltaRds} \ee We will not consider any detail of the lens
model, which can arise additional errors. From various sources of
the lens system, we can roughly say that the error of the velocity
dispersion is about $5$\% ($\Delta \sigma_v / \sigma_v \sim
5$\%)~\cite{SLACS}. That means we can distinguish the differences
between models if the difference of the ratio of the angular
diameter distances between models, $\Delta R_{ds} / R_{ds}$ is at
least $5$\%. Now we will compare the differences of the ratios of
angular diameter distances for different models. There was a work
related to this approach \cite{Sereno}. However in the previous
work, the equation of state parameter of dark energy was
constrained as a constant. We consider more general cases with the
stronger constraint on constant $\omega_{\rm{DE}}$s.

\subsection{$R_{ds}$ when $\omega_{\rm{DE}} =$ constant}

\begin{center}
\begin{figure}
\vspace{2cm} \centerline{ \psfig{file=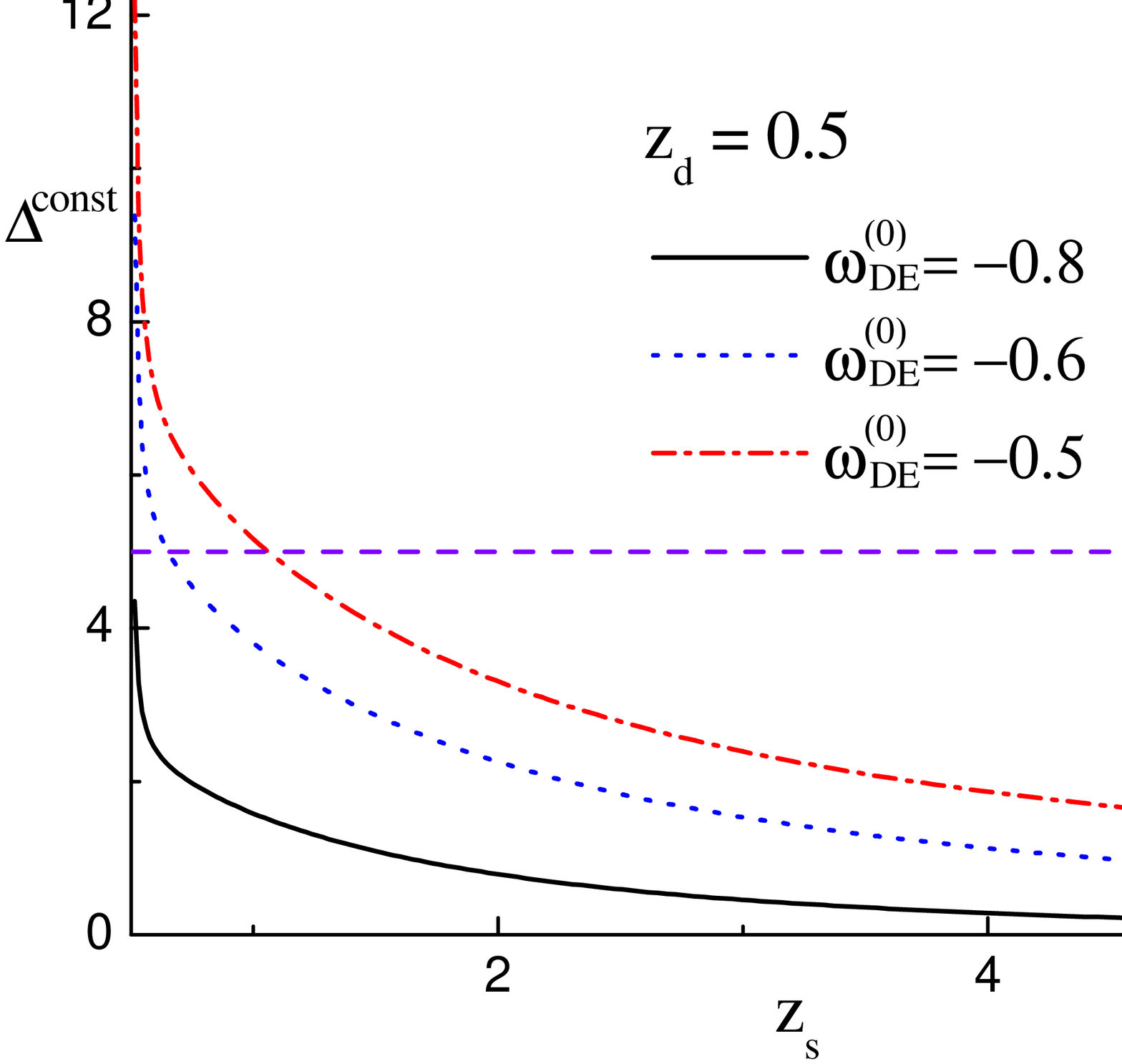,
width=8cm}\psfig{file=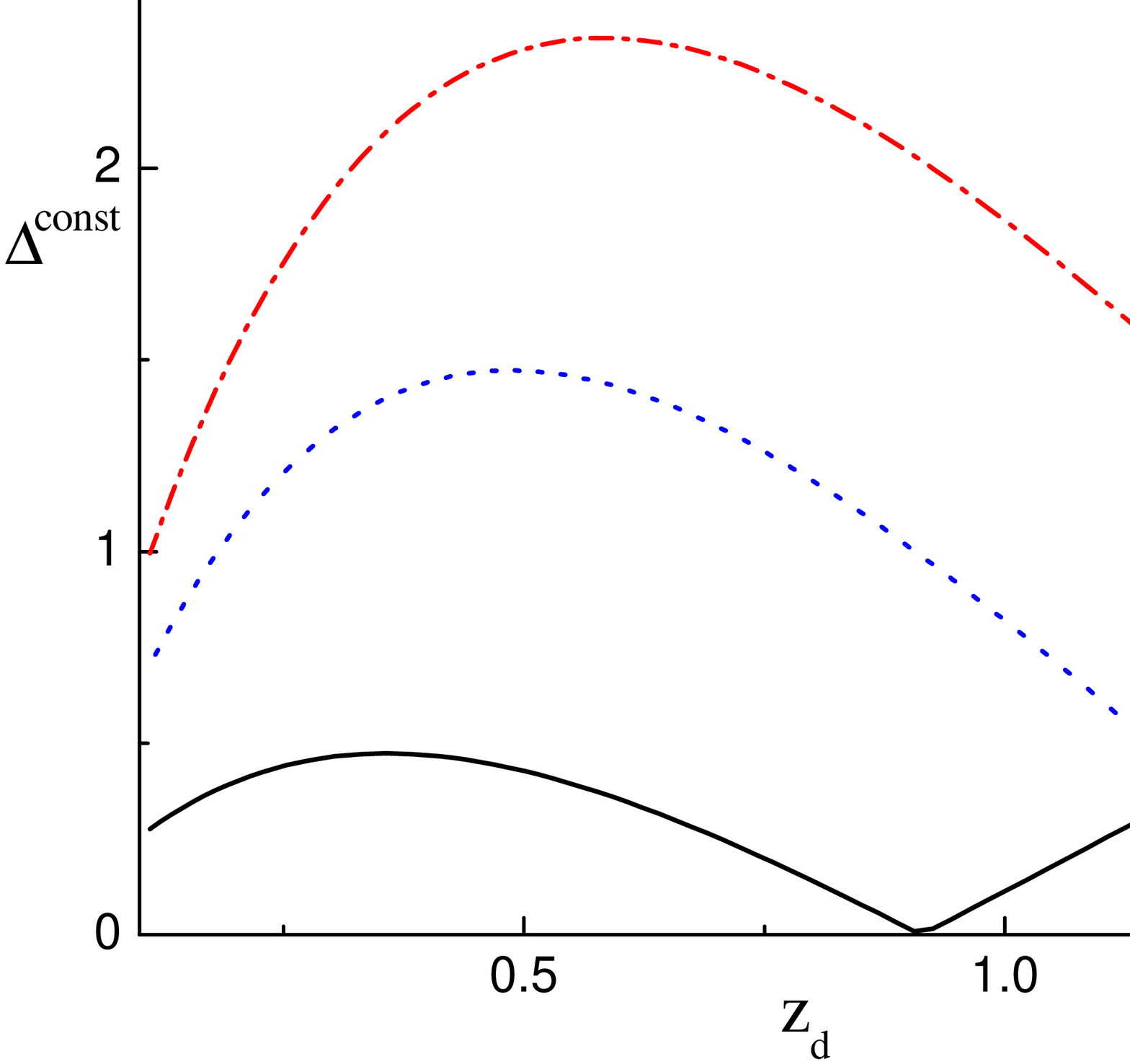, width=8cm} } \vspace{-2cm}
\caption{$\Delta^{\rm{const}}$ for the constant $\omega_{\rm{DE}}$
as a function of the location of the source ($z_s$). (a)
Dependence on the position of the source ($z_{s}$) of
$\Delta^{\rm{const}}$ for different values of
$\omega_{\rm{DE}}^{(0)}$, with the same notation for the label as
the left panel of Fig.~\ref{fig:ErrR} when $\Omega_{\rm{DE}}^{(0)}
= 0.73$. (b) Dependence on the position of lens ($z_{d}$) of
$\Delta^{\rm{const}}$ for different present values of
$\omega_{\rm{DE}}^{(0)}$, with the same notation as the left
panel.} \label{fig:ErrRzsd} \vspace{1cm}
\end{figure}
\end{center}

\vspace{-1.4cm}

We assume that the stand cosmological parameters are $\Ode = 0.73$
and $\odeo = -1.0$ and we denote the ratio of the angular diameter
distances $D_{ds}/D_{s}$ with these parameters as
$R_{ds}^{\Lambda}$. Figure~\ref{fig:ErrR} shows the difference of
$R_{ds}$ between the standard case and the various constant
equation of state cases, $\Delta^{\rm{const}} \equiv
|R_{ds}^{\Lambda} - R_{ds}^{\rm{const}}| / R_{ds}^{\Lambda} \times
100 (\%)$. In the left panel of Fig.~\ref{fig:ErrR}, we show the
$\Delta^{\rm{const}}$ dependence on $\Ode$ for different values of
$\odeo$, $-1.0$ (solid line), $-0.8$ (dotted line), and $-0.5$
(dash-dotted line). If there is an error in measuring $\Ode$
within the range $0.51 \leq \Ode \leq 0.87$~($0.52 \leq \Ode \leq
0.89$, $0.57 \leq \Ode $) for $\odeo = -1.0~(-0.8, -0.5$), then
$\Delta^{\rm{const}}$ is less than $5$\%. For example, if the
measured cosmological parameters are $\odeo = -0.5$ and $\Ode \geq
0.57$, then we cannot distinguish this with the standard case,
($\odeo = -1.0$ and $\Ode = 0.73$) within $5$\% by using $R_{ds}$,
{\it i.e.} by measuring $\theta_E$. $\Ode$ also varies for
different values of $\ode$, so $\odeo = -1.0$ does not necessarily
have smaller difference to the standard case as can be shown in
the left panel of Fig.~\ref{fig:ErrR}. When $\odeo = -1.0$, we can
have the standard case if $\Ode = 0.73$. In this case
$\Delta^{\rm{const}} = 0$. This is shown in the left panel of
Fig.~\ref{fig:ErrR} and it shows the consistency of figures. We
show $\Delta^{\rm{const}}$ dependence on $\odeo$ for different
values of $\Ode$, $0.85$ (solid line), $0.73$ (dotted line), and
$0.60$ (dash-dotted line) in the right panel of
Fig.~\ref{fig:ErrR}. Again if there is an error in measuring
$\odeo$ within the range $\odeo \leq -0.23$ ($-0.30$, $-0.44$) for
$\Ode = 0.85~(0.73, 0.60$), then $\Delta^{\rm{const}}$ is less
than $5$\%. For example, if the measured cosmological parameters
are $\Ode = 0.85$ and $\odeo \leq -0.23$, then we cannot
distinguish this with the standard case within $5$\%. We can also
see that $\Delta^{\rm{const}} = 0$ when $\Ode = 0.73$ and $\odeo =
-1.0$ in the right panel of Fig.~\ref{fig:ErrR}.

We show the dependence of $\Delta^{\rm{const}}$ on the position of
source ($z_s$) in the left panel of Fig.~\ref{fig:ErrRzsd} for
different values of $\odeo$ $-0.8$ (solid line), $-0.6$ (dotted
line), and $-0.5$ (dash-dotted line). We choose the position of
lens as $z_d = 0.5$ and $\Ode = 0.73$. If we choose $z_s = 3.0$
for different values of $\odeo$, then we have the same values of
$\Delta^{\rm{const}}$ of the dotted line of the right panel of
Fig.~\ref{fig:ErrR}. If the source position of lensing system is
located $z_s \geq 0.64$ ($1.04$) for $\odeo = -0.6$ ($-0.5$), then
$\Delta^{\rm{const}}$ is less than $5$\%. We cannot distinguish
the difference of $R_{ds}$s between the standard case and $\odeo =
-0.8$ within $5$\% independent of the source position. In the
right panel of Fig.~\ref{fig:ErrRzsd}, we show the dependence of
$\Delta^{\rm{const}}$ on the position of lens ($z_d$) when we
choose $z_s = 3.0$ and $\Ode = 0.73$ with the same notation as the
left panel of the figure for the different values of $\odeo$. When
we vary $z_d$ from $0.1$ to $1.5$ for every values of $\odeo$
($-0.8 \leq \odeo \leq -0.5$), $\Delta^{\rm{const}}$s are well
below $5$\%. To check the consistency between the figures, we can
check $\Delta^{\rm{const}}$ values when we choose $z_d = 0.5$ in
the right panel of Fig.~\ref{fig:ErrR}. This gives
$\Delta^{\rm{const}} \simeq 0.44$ ($1.5$, $2.3$) when $\odeo =
-0.8 (-0.6, -0.5$), which is equal to $\Delta^{\rm{const}}$ value
when we choose $z_s = 3.0$ in the left panel of the same figure.
\begin{center}
\begin{figure}
\vspace{2cm} \centerline{ \psfig{file=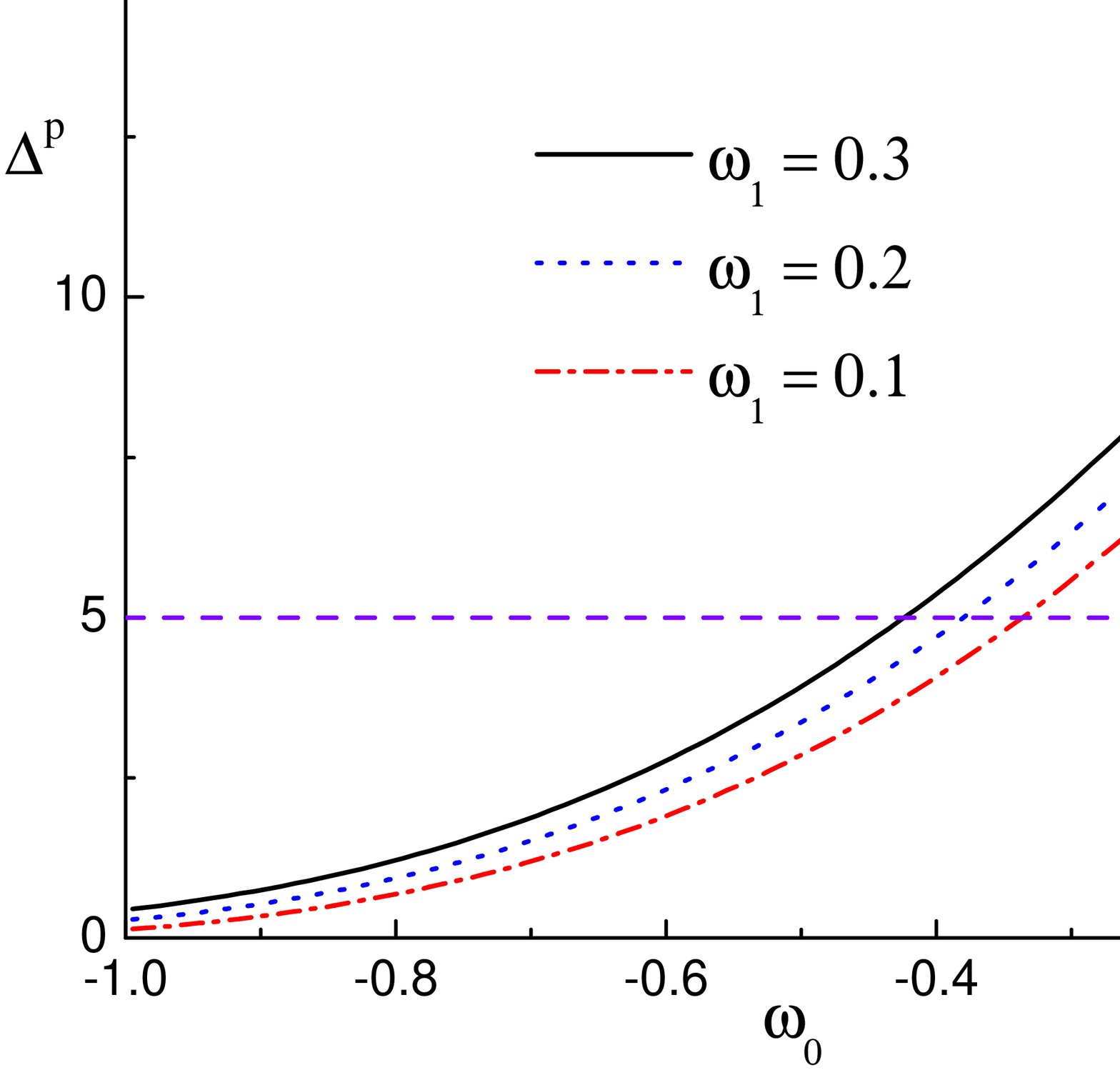,
width=8cm}\psfig{file=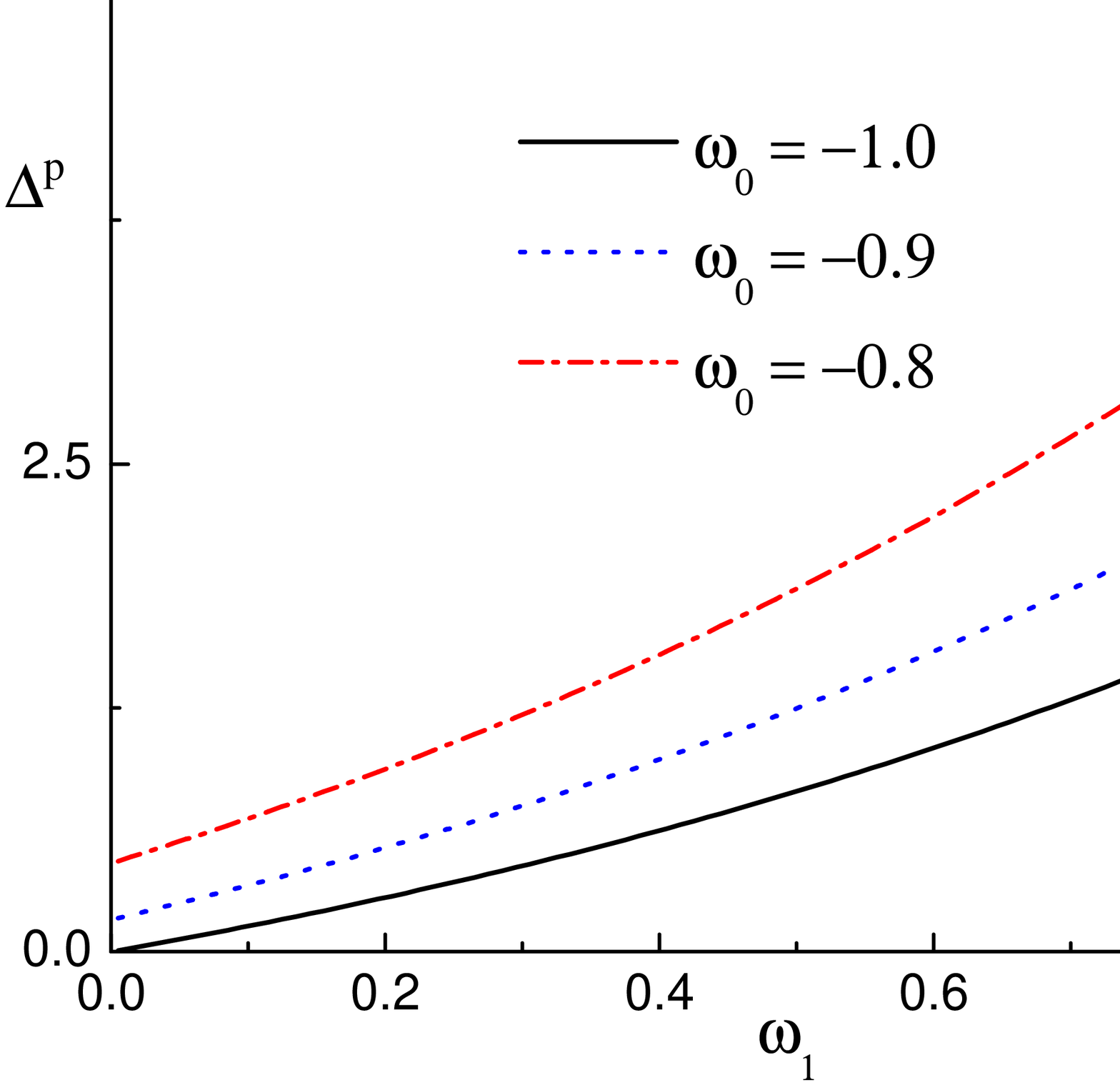, width=8cm} } \vspace{-2cm}
\caption{$\Delta^{\rm{p}}$ for the time varying equation of state
of dark energy ($\omega_{\rm{DE}} = w_0 + w_1(1-a)$). (a)
Dependence on the value of $\omega_0$ for different values of
$\omega_{1}$ : $0.1$ (solid line), $0.2$ (dotted line), and $0.3$
(dash-dotted line), respectively. (b) Dependence on the value of
$\omega_1$ of $\Delta^{\rm{p}}$ for different values of $\omega_0$
: $-1.0$ (solid line), $-0.9$ (dotted line), and $-0.8$
(dash-dotted line), respectively.} \label{fig:Errw} \vspace{1cm}
\end{figure}
\end{center}

\vspace{-1.5cm}

\subsection{$R_{ds}$ when $\omega_{DE} = \omega_0 + \omega_1 (1 - a)$}

Now we check the difference of the ratio of the angular diameter
distances $D_{ds}/D_{s}$ between the standard case and one of the
time varying $\ode$ models, $\ode = \omega_0 + \omega_1 (1-a)$
\cite{Linder}. We define $\Delta^{\rm{p}} \equiv |R_{ds}^{\Lambda}
- R_{ds}^{\rm{p}}|/R_{ds}^{\Lambda} \times 100 (\%)$ where
$R_{ds}^{\rm{p}}$ means $R_{ds}$ value when we use $\ode =
\omega_0 + \omega_1 (1-a)$. We assume that there is no error in
the measured value of $\Ode = 0.73$ in this case. We show
$\Delta^{\rm{p}}$ dependence on $\omega_0$ value for different
values of $\omega_1$, $0.3$ (solid line), $0.2$ (dotted line), and
$0.1$ (dash-dotted line), respectively in the left panel of
Fig.~\ref{fig:Errw}. $\Delta^{\rm{p}}$ is within $5$\% for
$\omega_1 = 0.3$ ($0.2, 0.1$) when $\omega_0  \leq -0.43$ ($-0.38,
-0.34$). For example, when $\omega_1 = 0.3$ and $\omega_0 \leq
-0.43$, then we cannot distinguish the cosmology with the
cosmological constant from the cosmological model with time
varying dark energy $\ode = \omega_0 + \omega_1 (1 - a)$ within
$5$\% by using $R_{ds}$. In the right panel of
Fig.~\ref{fig:Errw}, we show the $\omega_1$ dependence of
$\Delta^{\rm{p}}$ for different values of $\omega_0$, $-1.0$
(solid line), $-0.9$ (dotted line), and $-0.8$ (dash-dotted line).
When $\omega_1$ varies from $0$ to $1$ every $\Delta^{\rm{p}}$
value is within $5$\% for every value of $\omega_0$ from $-1.0$ to
$-0.8$. For the consistency check we can choose $\omega_1 = 0.3$
for three different lines. We can check these values with the
three different points of solid line of the left panel of the
figure. When we choose $\omega_1 = 0.3$, $\Delta^{\rm{p}} = 0.45
0.76$, and $1.2$ for $\omega_0 = -1.0, -0.9$, and $-0.8$ in both
left and right panels of Fig.~\ref{fig:Errw}.
\begin{center}
\begin{figure}
\vspace{2cm} \centerline{ \psfig{file=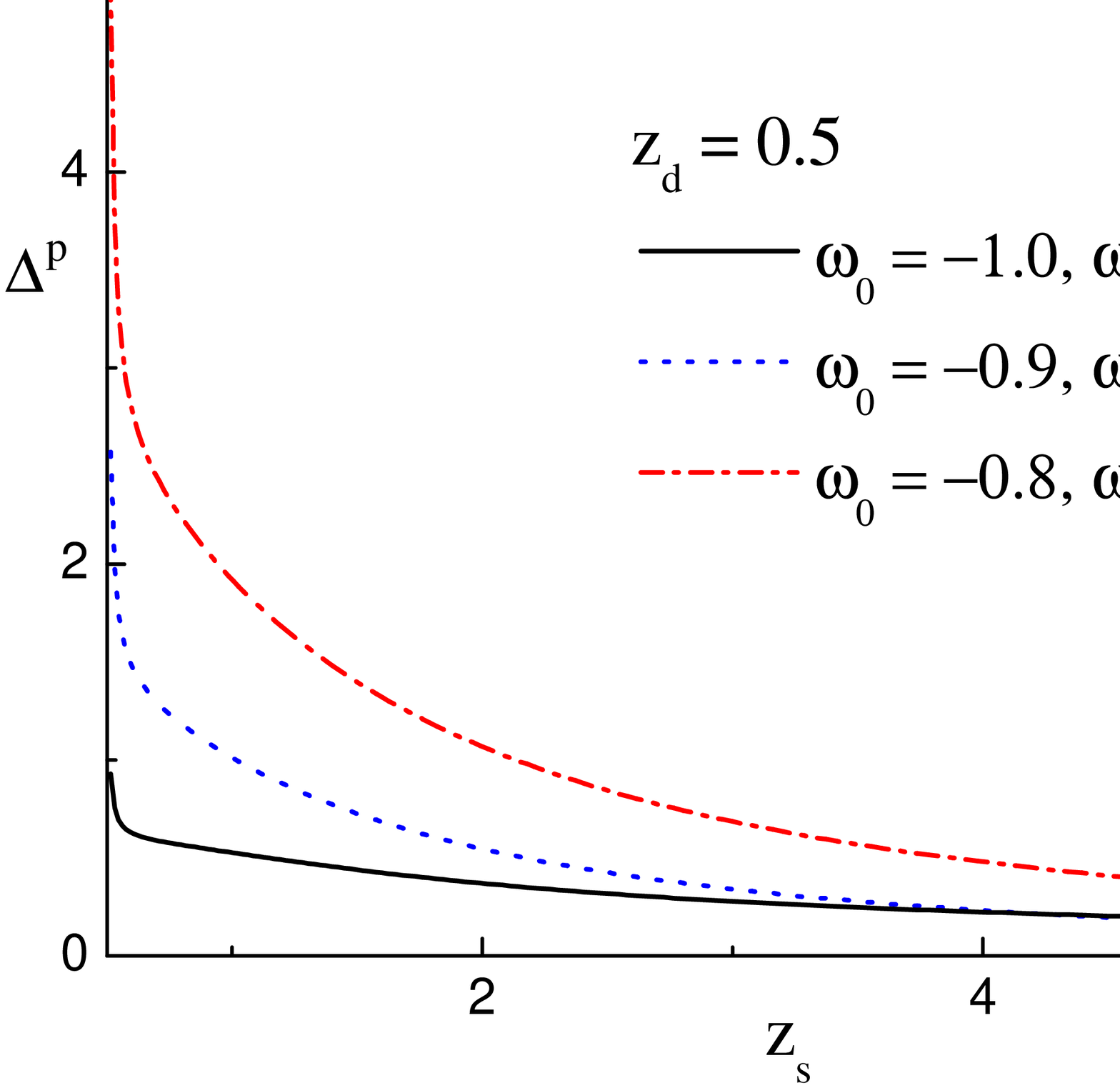,
width=8cm}\psfig{file=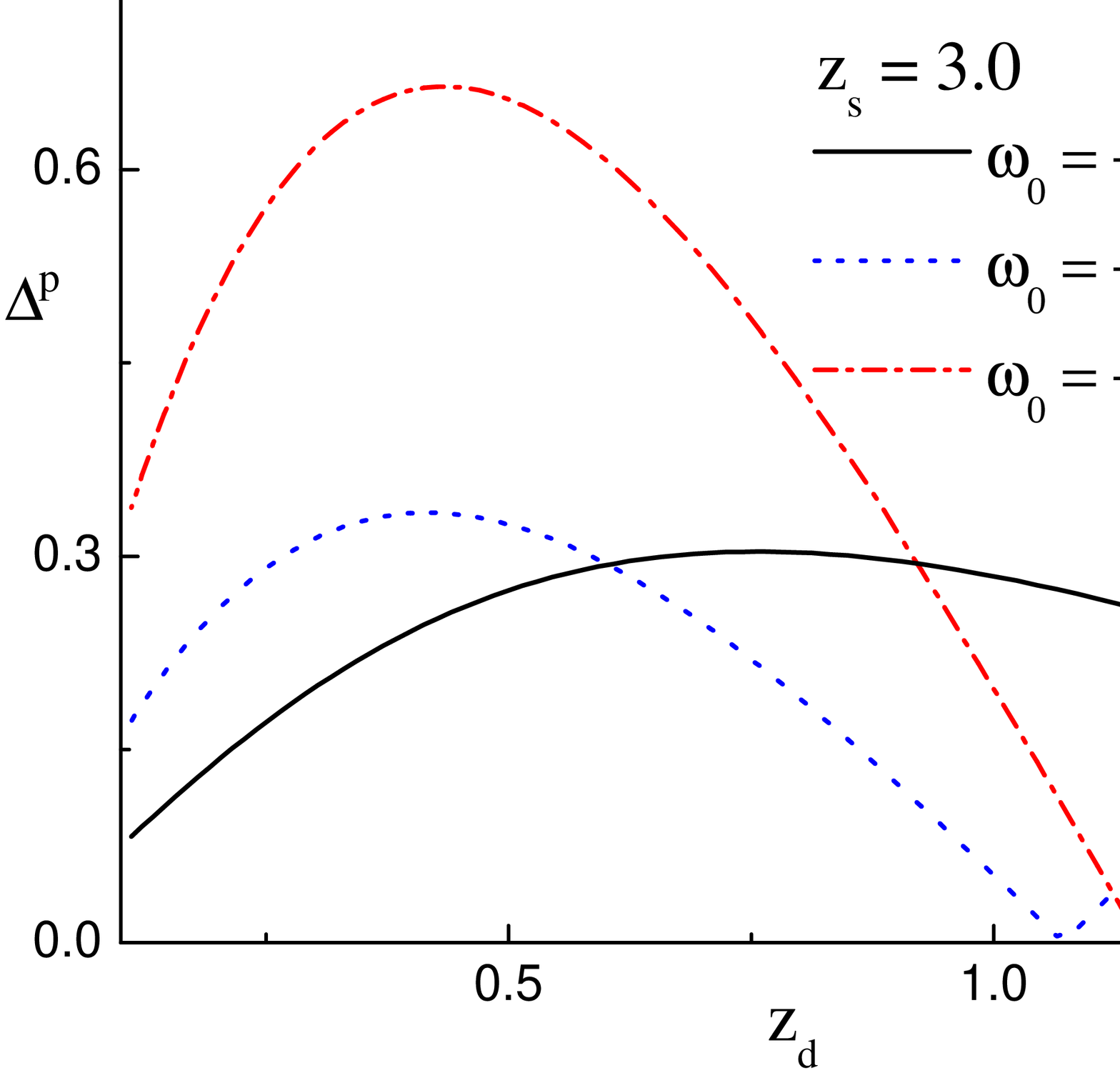, width=8cm} } \vspace{-2cm}
\caption{$\Delta^{\rm{p}}$. (a) Dependence on $z_{s}$ of
$\Delta^{\rm{p}}$ for various sets of $\omega_{0}$ and $\omega_1$.
(b) Dependence on $z_{d}$ of $\Delta^{\rm{p}}$ for the same sets
of $\omega_{0}$ and $\omega_1$ as in the left panel.}
\label{fig:Errwoz} \vspace{1cm}
\end{figure}
\end{center}

\vspace{-1.5cm} We again check $\Delta^{\rm{p}}$ dependence on the
position of $z_s$ and $z_d$ for the various sets of $\omega_0$ and
$\omega_1$ values in figure \ref{fig:Errwoz}. In the left panel of
Fig.~\ref{fig:Errwoz}, we show $\Delta^{\rm{p}}$ dependence on
$z_s$. We choose $z_d = 0.5$ and $\Ode = 0.73$ in this case. For
the given sets of $\omega_0$ and $\omega_1$ values, we cannot
distinguish the cosmological model with the cosmological constant
from the cosmological model with time varying dark energy $\ode =
\omega_0 + \omega_1 (1 - a)$ no matter where the source of lens is
located from $z_s = 0.5$ to $6$ within $5$\% by measuring
$R_{ds}$. We check $z_d$ dependence on $\Delta^{\rm{p}}$ for the
same sets of $\omega_0$ and $\omega_1$ in the right panel of
Fig.~\ref{fig:Errwoz}. We choose $z_s = 3.0$. Again none of case
can be distinguished with cosmological constant case within $5$\%.
\begin{center}
\begin{figure}
\vspace{2cm} \centerline{ \psfig{file=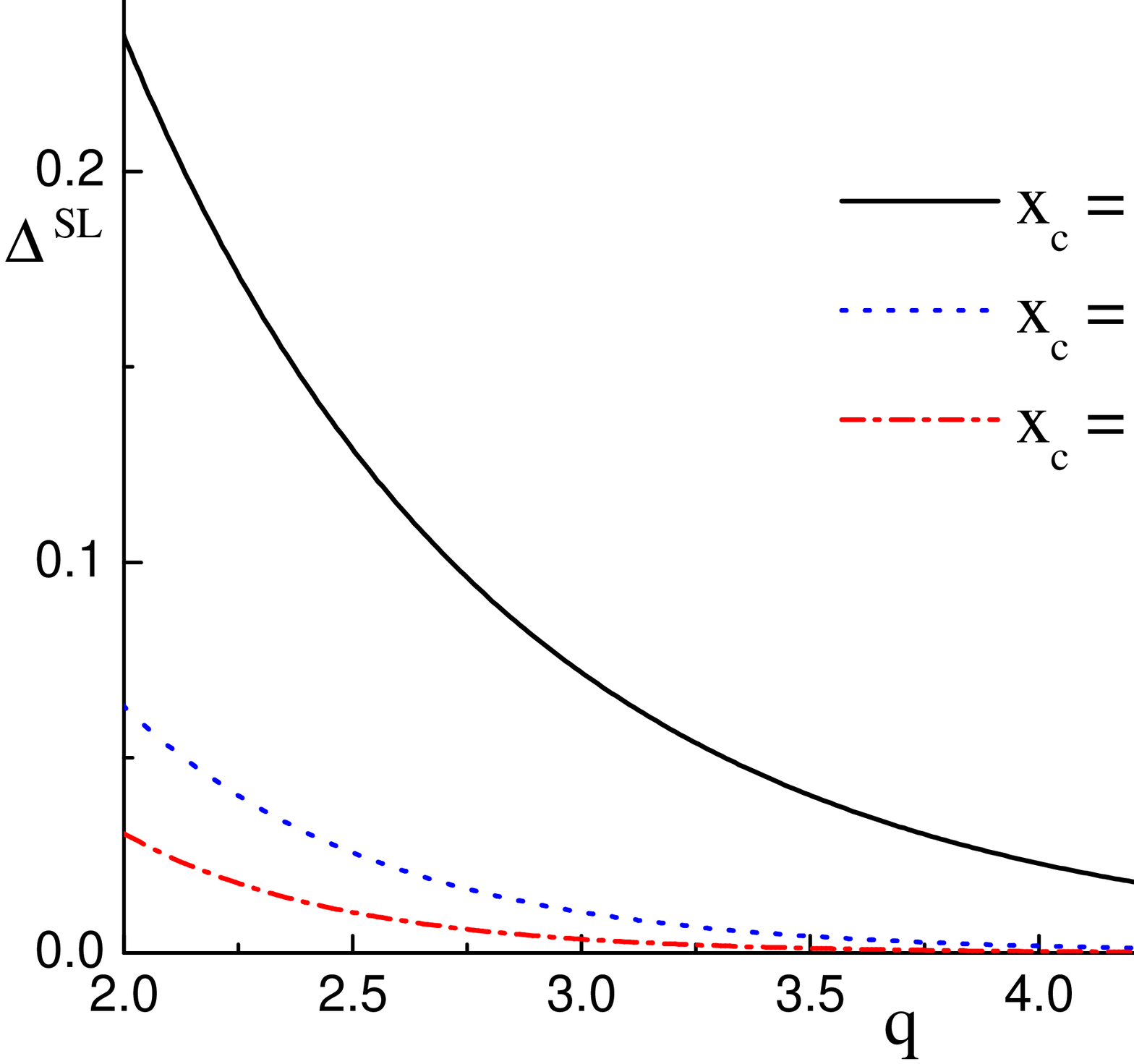,
width=8cm}\psfig{file=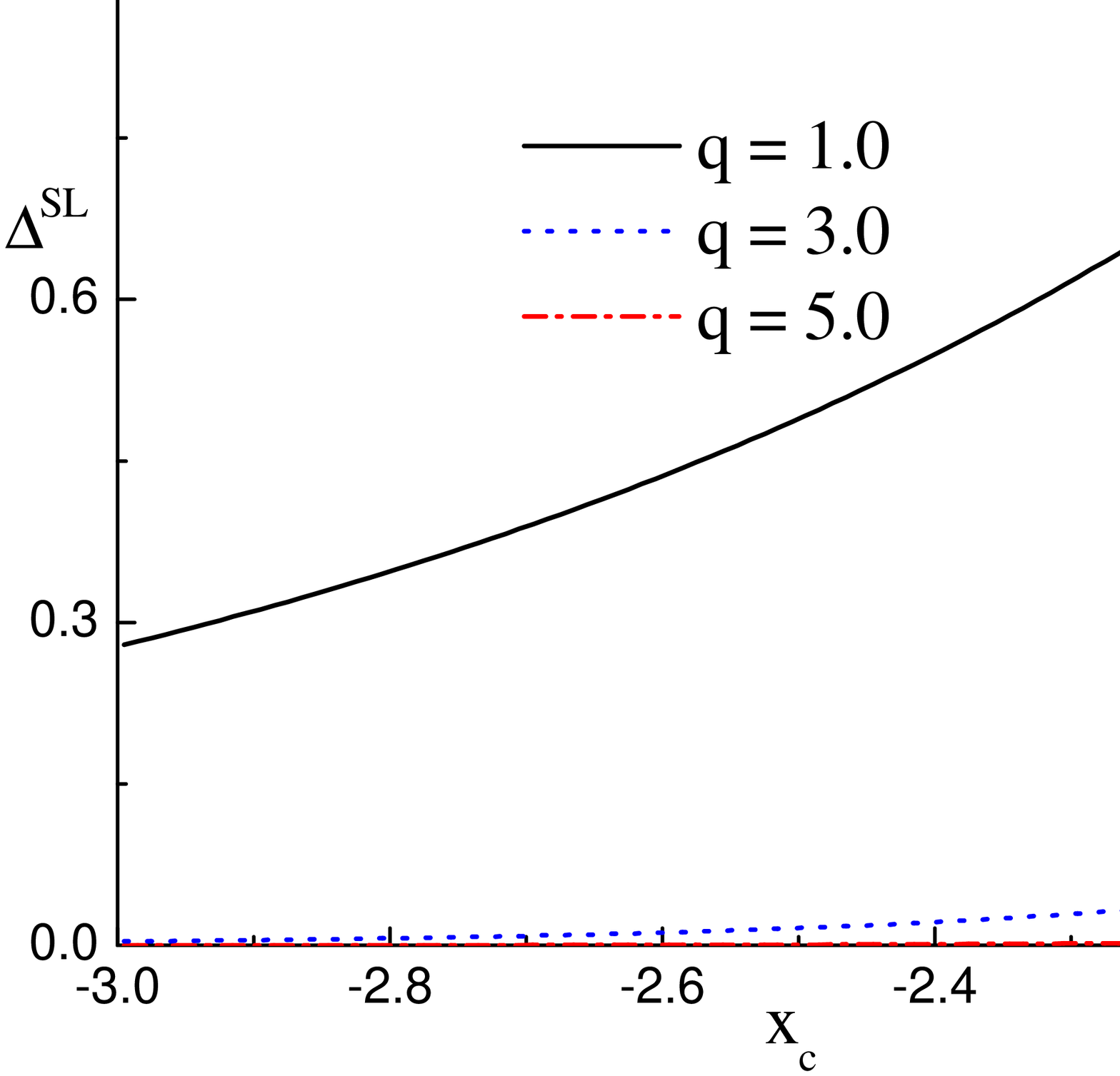, width=8cm} } \vspace{-2cm}
\caption{$\Delta^{\rm{SL}}$ for the varying equation of state of
dark energy ($\omega_{\rm{DE}} = -1 + 4 a_c^q / 3(a^q + a_c^q) $).
(a) Dependence on $q$ of $\Delta^{\rm{SL}}$ for different values
of $x_c$ : $-2.00$ (solid line), $-2.64$ (dotted line), and
$-3.00$ (dash-dotted line), respectively. (b) Dependence on $x_c$
of $\Delta^{\rm{SL}}$ for different values of $q$ : $1.0$ (solid
line), $3.0$ (dotted line), and $5.0$ (dash-dotted line),
respectively.} \label{fig:Errwp} \vspace{1cm}
\end{figure}
\end{center}

\vspace{-1.0cm}

\subsection{$R_{ds}$ when $\omega_{\rm{DE}} = -1 + 4a_c^q / 3(a^q +
a_c^q)$}

We check another time varying dark energy model by using the
parameter $\ode = -1 + 4 a_c^q / 3(a^q + a_c^q)$ in figure
\ref{fig:Errwp}. We define $\Delta^{\rm{SL}} \equiv
|R_{ds}^{\Lambda} - R_{ds}^{\rm{SL}}| / R_{ds}^{\Lambda} \times
100 (\%)$ where $R_{ds}^{\rm{SL}}$ means $R_{ds}$ value when we
use $\ode = -1 + 4 a_c^q / 3(a^q + a_c^q)$ \cite{SLee2}. The
angular diameter distance of this parametrization is almost
degenerate with that of the cosmological constant. Thus, the ratio
difference of angular diameter distances between this
parametrization and the cosmological constant is extremely small
for the different choice of parameters. We choose $\Ode = 0.73$,
$z_s = 3.0$, and $z_d = 0.5$ in this figure. In the left panel of
Fig.~\ref{fig:Errwp}, we show the dependence of $\Delta^{\rm{SL}}$
on $q$ for the different values of $x_c$, $-2.00$ (solid line),
$-2.64$ (dotted line), and $-3.00$ (dash-dotted line). For all of
the cases, $\Delta^{\rm{SL}}$ are within $1$\% for different
values of $x_c$ from $q = 2$ to $5$. In the right panel of
Fig.~\ref{fig:Errwp}, we show $\Delta^{\rm{SL}}$ dependence on
$x_c$ for different values of $q$, $1$ (solid line), $3$ (dotted
line), and $5$ (dash-dotted line). Except for $q=1$ case,
$\Delta^{\rm{SL}}$ are degenerate within $1$\%. Thus, we cannot
tell any difference between the cosmological constant and the dark
energy with the parameter $\ode = -1 + 4 a_c^q / 3(a^q + a_c^q)$
by measuring $R_{ds}$ within $5$\%.

\begin{center}
\begin{figure}
\vspace{2cm} \centerline{ \psfig{file=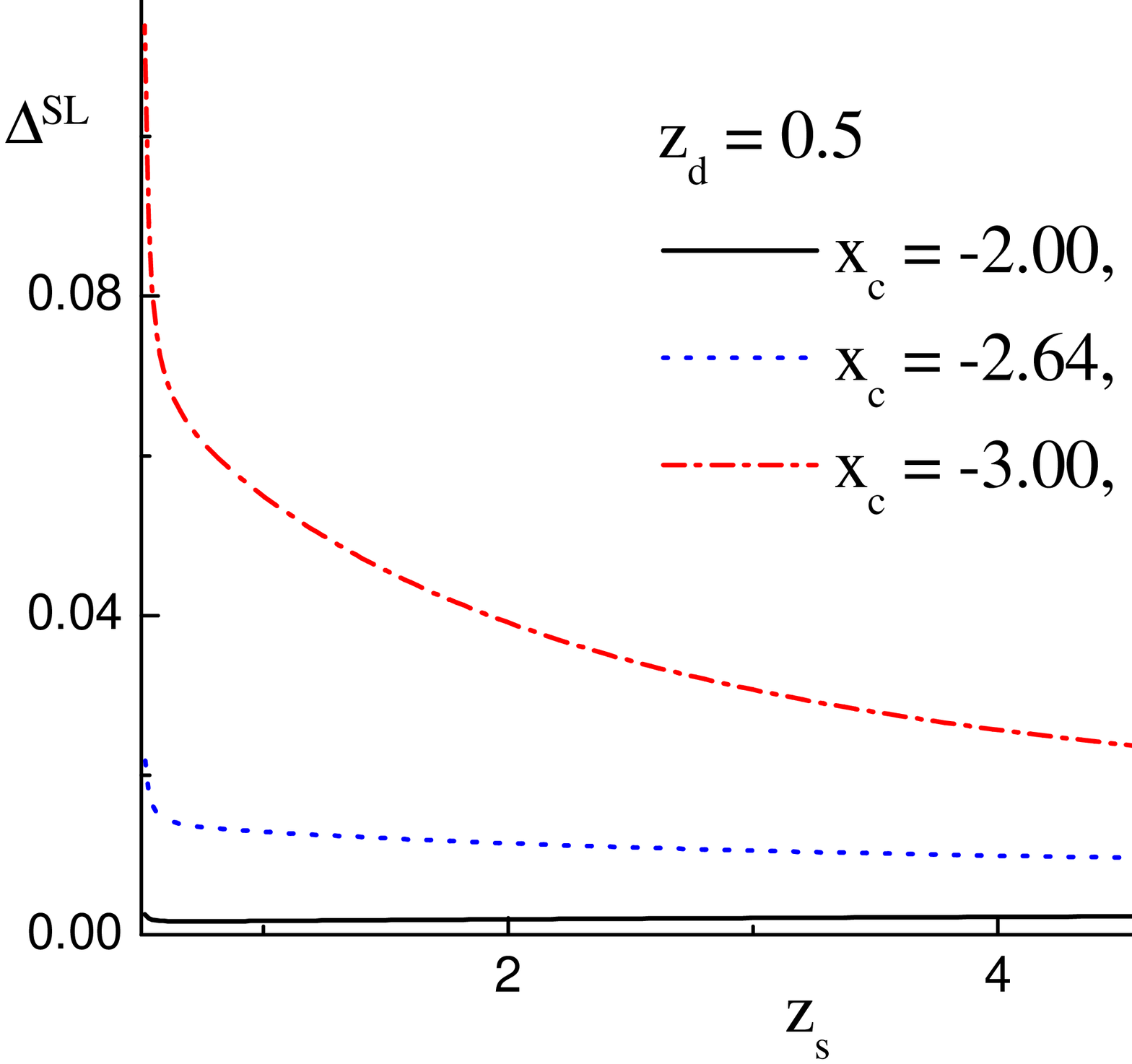,
width=8cm}\psfig{file=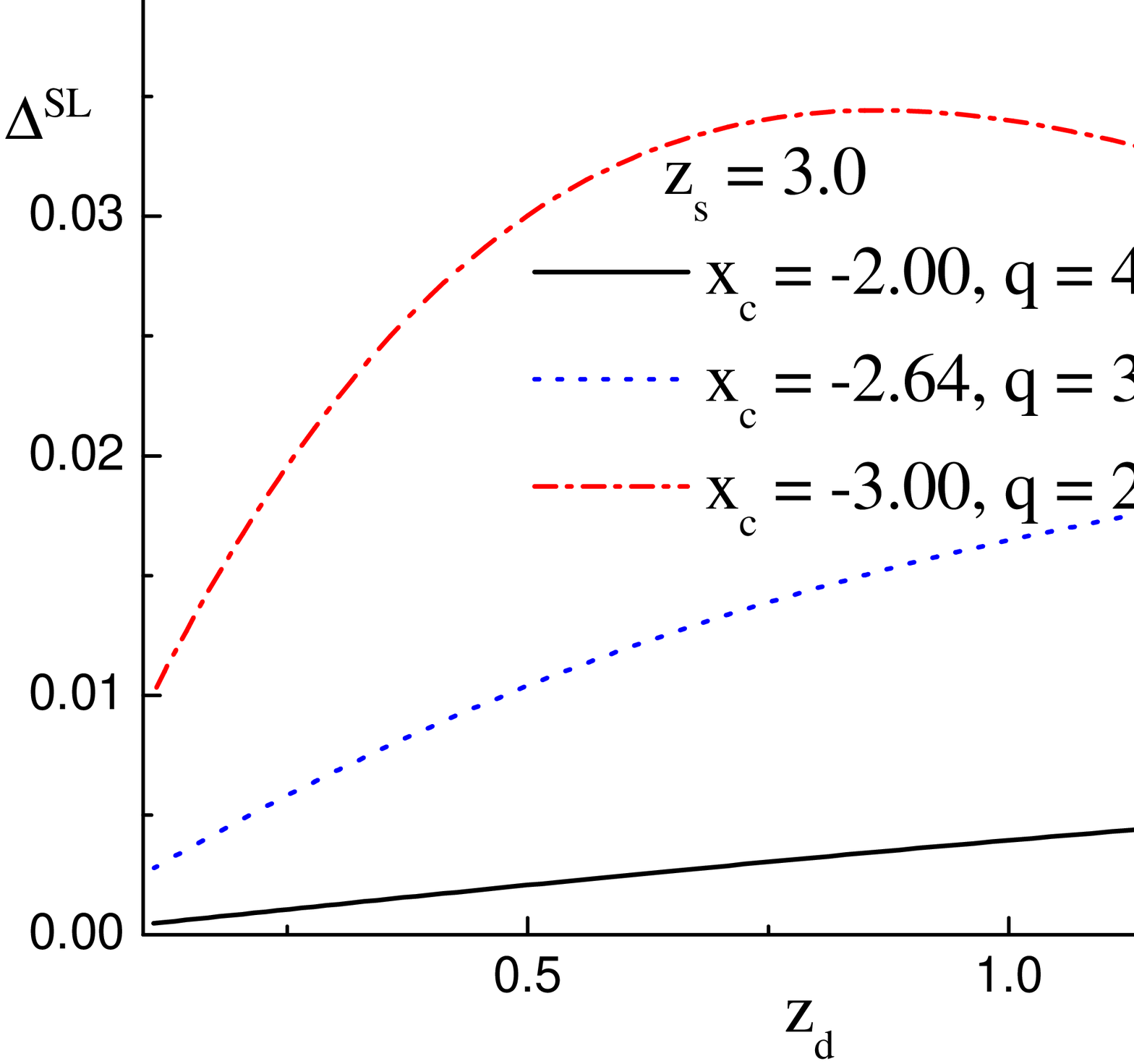, width=8cm} } \vspace{-2cm}
\caption{$\Delta^{\rm{SL}}$. (a) Dependence on $z_{s}$ of
$\Delta^{\rm{SL}}$ for various sets of $x_c$ and $q$ values. (b)
Dependence on $z_{d}$ of $\Delta^{\rm{SL}}$ for the same sets of
$x_c$ and $q$ values as in the left panel.} \label{fig:Errwpz}
\vspace{1cm}
\end{figure}
\end{center}

We again check $\Delta^{\rm{SL}}$ dependence on the position of
$z_s$ and $z_d$ for various sets of $x_c$ and $q$ values in
Fig.~\ref{fig:Errwpz}. In the left panel of Fig.~\ref{fig:Errwpz},
we show $\Delta^{\rm{SL}}$ dependence on $z_s$. We choose $z_d =
0.5$ and $\Ode = 0.73$ in this case. For given sets of ($x_c$,
$q$) values, ($-2.00$, $q =4$) (solid line), ($-2.64$, $q = 3$)
(dotted line), and ($ -3.00$, $q =2$) (dash-dotted line), we
cannot distinguish the cosmological model with the cosmological
constant from the cosmological model with time varying dark energy
$\ode = -1 + 4 a_c^q /3(a^q +a_c^q)$ no matter where the source of
lens is located from $z_s = 0.5$ to $6$ within $1$\% by measuring
$R_{ds}$. We check $z_d$ dependence on $\Delta^{\rm{SL}}$ for the
same sets of $x_c$ and $q$ in the left panel of the figure. We
choose $z_s = 3.0$. Again none of the cases can be distinguished
from the cosmological constant case within $1$\%.

\vspace{-1.0cm}
\section{Conclusions}
\setcounter{equation}{0}
We have investigated the angular diameter distances from the
source to the present observer for the various dark energy models.
The angular diameter distance is proportional to the luminosity
distance, which is used for the analysis of Type Ia supernovae.
The difference of $H_0D_s$ between models is apparent and
measuring this quantity as to probe dark energy models might be
proper for further investigation. In addition to SNeIa, the
gravitational lensing has been known as the possible method to
probe the property of dark energy.

Most of the lensing analyses have the Einstein radius as a basic
quantity. This is lens model independent observable quantity and
might be a good probe for the property of dark energy. The
Einstein radius is proportional to both the ratio of the angular
diameter distances and the velocity dispersion squared. However,
we have shown that there is degeneracy between different models
for the value of the ratio of the angular diameter distances. If
the error in measuring the velocity dispersion exceeds the
difference of the ratio of the angular diameter distances of two
different dark energy models, then we cannot distinguish the
differences between different dark energy models by measuring the
Einstein radius. In our analysis we have shown that this is the
case for most of the parameter spaces of the dark energy models.
Thus, a single strong gravitational lensing might not be a proper
method to investigate the property of dark energy. However, better
understanding to the mass profile of clusters in the future or
other methods related to arc statistics rather than the distances
may be still used for constraints on dark energy.

\section{Acknowledgements}
\setcounter{equation}{0}
S.L. would like to thank K. Umetsu for fruitful discussion and
thank S-Y. Tsai for useful help in the compilation of this
manuscript. K-W.N. was supported in part by the National Science
Council, Taiwan, ROC under the grant NSC94-2112-M-001-024.


\end{document}